\pdfoutput=1
\RequirePackage[T1]{fontenc}
\documentclass[12pt]{article}

\usepackage[height=8.85in,width=6.45in]{geometry}

\usepackage[utf8]{inputenc}
\usepackage{amsmath}
\usepackage{amssymb}
\usepackage{mathtools}
\usepackage{dsfont}
\numberwithin{equation}{section}
\usepackage{slashed}
\usepackage{braket}
\usepackage{enumitem}
\usepackage{tabu}
\usepackage[svgnames,psnames]{xcolor}
\usepackage[colorlinks,citecolor=DarkGreen,linkcolor=FireBrick,urlcolor=FireBrick,linktocpage,unicode,psdextra]{hyperref}
\usepackage{cite}
\usepackage{comment}
\usepackage{graphicx}
\usepackage{tikz}
\usepackage{tikz-cd}
\usetikzlibrary{calc,topaths,decorations,decorations.pathmorphing,arrows,decorations.markings,cd}
\tikzset{
->-/.style args={#1rotate#2}{decoration={markings, mark=at position #1 with {\arrow[scale=1.5,rotate = #2 ]{stealth}}}, postaction={decorate}}
}

\tikzset{line/.style={line width=0.25mm},
curve/.style={line,smooth,tension=1},
->-/.style={decoration={
  markings,
  mark=at position #1 with {\arrow[>=stealth]{>}}},postaction={decorate}},
-<-/.style={decoration={
  markings,
  mark=at position #1 with {\arrow[>=stealth]{<}}},postaction={decorate}},
}

\usepackage[bottom]{footmisc}

\usepackage{times}
\usepackage{courier}
\usepackage{bm}
\usepackage{subfig}
\usepackage{tcolorbox} 

\usepackage{mdframed}

\renewenvironment{figure}[1][]{
  \begin{originalfigure}[#1]
    \begin{mdframed}[linecolor=black!0,backgroundcolor=black!1]
}{
    \end{mdframed}
  \end{originalfigure}
}

\def\cH{\mathcal{H}}

\def\D{\mathbb{D}}

\def\cH{\mathcal{H}}
\def\cA{\mathcal{A}}
\def\cC{\mathcal{C}}

\def\barray{\begin{eqnarray}}
\def\earray{\end{eqnarray}}
\def\beq{\begin{equation}}
\def\eeq{\end{equation}}

\def\C{{\cal C}}
\def\N{{\cal N}}
\def\D{{\cal D}}

\def\a{\mathsf{a}}
\def\b{\mathsf{b}}
\def\c{\mathsf{c}}
\def\d{\mathsf{d}}
\def\x{\mathsf{x}}

\def\aa{\mathfrak{a}}

\newcommand{\id}{\mathds{1}}

\IfFileExists{ying}{
  \usetikzlibrary{external}
  \tikzexternalize[prefix=./figures/]
}

\usetikzlibrary{calc,shapes.geometric}
\pgfmathparse{atan2(0,1)}
\ifdim\pgfmathresult pt=0pt
  \tikzset{declare function={atanXY(\x,\y)=atan2(\y,\x);atanYX(\y,\x)=atan2(\y,\x);}}
\else                 
  \tikzset{declare function={atanXY(\x,\y)=atan2(\x,\y);atanYX(\y,\x)=atan2(\x,\y);}}
\fi

\begin{document}

\begin{titlepage}


\vskip 3cm

\begin{center}

{\Large \bfseries  
Entropic Order Parameters for Categorical Symmetries in 2D-CFT}
\vskip 1 cm

\vskip 1cm
J. Molina-Vilaplana$^1$,
P. Saura-Bastida$^1$,
and G. Sierra$^2$,
\vskip 1cm
\begin{tabular}{ll}
1 & Universidad Politécnica de Cartagena, Cartagena, Spain\\
2 & Instituto de Física Teórica, UAM/CSIC, Universidad Autónoma de Madrid, Madrid, Spain
\end{tabular}
\end{center}

\noindent 
{\bf Abstract:} In this work, we propose an information theoretic order parameter able to characterize the presence and breaking of categorical symmetries in $(1+1)$-d  rational conformal field theories (RCFT). Specifically, we compute the quantum relative entropy between the ground states of RCFTs representing the critical point of phase transitions between different symmetry-broken phases of theories with categorical symmetries, and their \emph{symmetrized} versions. We find that, at leading order in the high temperature limit, this relative entropy only depends on the expectation values of the quantum dimensions of the topological operators implementing the categorical symmetry. This dependence suggests  that our proposal can be used to characterize the different broken phases of  $(1+1)$-d theories with categorical symmetries.
\vspace{1 cm}

\noindent 
{\bf Keywords:} Relative Entropy, Categorical Symmetries, Conformal Field Theory, Symmetry Breaking

\end{titlepage}
\setcounter{tocdepth}{2}
\tableofcontents

\section{Introduction}
\label{sec:introduction}
The last few years have witnessed  remarkable developments in the notion of a global symmetry in theoretical physics.  The major idea comes from realizing that every symmetry operation can be associated with a topological operator \cite{Gaiotto:2014}. In this new framework,  symmetries are not only captured by a discrete or continuous group, but also by more general mathematical structures such as higher groups and fusion categories.  This new viewpoint, known as \emph{categorical symmetry}, has led to a paradigm shift and fostered new directions in the study of quantum field theories (QFT) (see \cite{McGreevy:2022} for a comprehensive and pedagogical review on these developments).

Global symmetries are central to the understanding of physics at low energies. As a fact, they characterize the different phases of a theory in terms of spontaneously symmetry-broken (SSB) phases and the corresponding  phase transitions between such phases. For group-like symmetries, this refers to what is known as the Landau paradigm. Due to the extension of the concept of global symmetry represented by categorical symmetries, there has been a recent interest in the study of phases in QFT constrained by these categorical symmetries \cite{Bhardwaj:2023, Bhardwaj:2023b}. The aim is  thus to generalize the standard Landau theory to a \emph{categorical Landau paradigm} in which novel phases and phase transitions may arise.
In this scenario, for instance, it has been shown that for (1+1)d gapped phases with categorical symmetries, contrary to the case of the usual group-like symmetries,  the different vacua of a gapped phase may be physically distinguishable  \cite{Bhardwaj:2023}. This depends on the properties of interfaces between two distinct vacua which are given by the topological line defects of the non-invertible symmetry. 

 In order to provide new tools to illuminate the role of generalized categorical symmetries in QFT, here, following \cite{Casini:2020}, we propose an order parameter able to characterize the presence and breaking of these symmetries and, thus, the new phases of matter related to those symmetries. This order parameter can be naturally defined using information theoretic quantities and thus can be cast as an entropic order parameter. 

Specifically, we study a concrete class of $(1+1)$-d QFT known as rational conformal field theories (RCFT), representing the critical point of phase transitions between the different $(1+1)$-d symmetry-broken phases of theories with categorical symmetries, and the entropic order parameter proposed here allows us to focus on the vacuum states of these theories, which are defined on a flat Euclidean spacetime. Following \cite{Casini:2020}, we use the information theoretic quantity known as the quantum relative entropy, which intuitively, quantifies the distinguishability between two quantum states \cite{Petz:2004}. It is known that the use of relative entropy in the context of QFT poses several advantages with respect to other entropic quantities, such as the entropy \cite{Casini:2020, Casini:2019, Magan:2020}. Namely, all other information quantities can be
derived from it.

The paper is structured as follows: In Section \ref{sec:background} we present a brief overview of the basic material covering RCFT and topological operators implementing global categorical symmetries. In Section \ref{sec:averages} we  introduce a generalization of the concept of \emph{symmetrization} of quantum states for the case of modular fusion categories (MFC) in RCFT. Then, following \cite{Casini:2020}, the relative entropy between density matrices and their \emph{symmetrizations} over a MFC, is proposed as an entropic order parameter. In Section \ref{sec:rel_ent_cat} we compute the relative entropy between the ground states of RCFTs representing the critical point of phase transitions between different symmetry-broken phases of theories with categorical symmetries and their \emph{symmetrized} versions. We find that at leading order, this relative entropy only depends on the expectation values of the TDLs or quantum dimensions of the topological operators implementing the symmetry. Finally, we present our conclusions and perspectives for extensions of this work in Section \ref{sec:conclusions}.

\section{Background}
\label{sec:background}
In this section, we review some very basic facts about RCFT, and topological operators implementing categorical symmetries in these theories. A more comprehensive treatment of these topics can be found in \cite{Chang:2018}.

We consider diagonal RCFTs with a chiral algebra $\cA$ and its modular tensor category (MTC) $\cC$ given in terms of  simple objects $\a,\b,\c,\ldots \in \cC$. 

In these theories, topological lines are in one-to-one correspondence with its bulk primaries and are called Verlinde lines. Those  act on the bulk primaries as \cite{Chang:2018},
\begin{align}
    \b |\phi_{a}\rangle = \frac{S_{\b\a}}{S_{1\a}}|\phi_{a}\rangle, \label{tdl_local}
\end{align}
where $S_{\a\b}$ denotes an element of the unitary $S$-matrix of the RCFT.

The TDLs $\a \in \cC$, satisfy the same fusion algebra as that of the bulk primaries, that is
\begin{align}
    \a \times\b = \sum_{c \in \cC} N^{\c}_{\a\b} \c \,, \label{fuse_l}
\end{align}
where $N^{\c}_{\a\b}\in\mathbb{Z}_{\geq 0}$ are the fusion coefficients given by the Verlinde formula \cite{Verlinde:1988},
\beq\label{eq:verlinde}
N^{\c}_{\a\b} = \sum\limits_{\d \in \cC}\frac{S_{\d \a}S_{\d \b}S^{*}_{\d\c}}{S_{1\d}}\, .
\eeq

An important property of any TDL $\b \in \C$ is its {\it quantum dimension} $d_\b$. This is defined as the vacuum expectation value,
\begin{align}
    \langle \id | \b |\id \rangle = \frac{S_{\b\id}}{S_{\id \id}} \equiv d_\b\, \label{q_dim}
\end{align}
In any unitary CFT with a unique vacuum, $d_\b\geq 0$\cite{Chang:2018}.

We consider the torus partition function (with the temporal $S^1$ along the vertical axis with length $T$ and the spatial $S^1$ along the horizontal axis with length $L$)  given by ,
\begin{align}
    Z[q,\bar{q}] = {\rm Tr}\left(q^{L_0-\frac{c}{24}}\bar{q}^{\bar{L}_0-\frac{\bar{c}}{24}}\right) = \raisebox{-11pt}{
\begin{tikzpicture}
\draw [dashed, decoration = {markings, mark=at position 0.6 with {\arrow[scale=1]{stealth}}}, postaction=decorate] (-1,0.5) -- node[below]{$_\mathds{1}$} (1,0.5) ;
\draw (-1,0) rectangle (1,1);
\end{tikzpicture}
} \label{partn_fn}
\end{align}
where $q=e^{2\pi\text{i}\tau}$, and $\tau=iT/L$, the modular parameter on the torus, so, $q=e^{-2\pi(T/L)}$.
The density matrix $\rho$ of the ground state is given thus by 
\beq
\rho = \frac{q^{L_0-\frac{c}{24}}\bar{q}^{\bar{L}_0-\frac{\bar{c}}{24}}}{Z[q,\bar{q}]}\, .
\label{dens_mat}
\eeq

In this work, we will consider insertions of TDLs $\a$ along the spatial $S^1$ which can be depicted as:
\begin{align}
    Z[q,\bar{q}, \a] = {\rm Tr}\left(\a \, q^{L_0-\frac{c}{24}}\bar{q}^{\bar{L}_0-\frac{\bar{c}}{24}}\right) = \raisebox{-11pt}{
\begin{tikzpicture}
\draw [red, thick, decoration = {markings, mark=at position 0.6 with {\arrow[scale=1]{stealth}}}, postaction=decorate] (-1,0.5) -- node[below] {$\a$} (1,0.5);
\draw (-1,0) rectangle (1,1);
\end{tikzpicture}
}
\label{partn_fn_spatial}
\end{align}
which, from here in advance we simply denote by $Z[q,\a]$. It is important to note that,
\begin{align}
    Z[q,\a] &= 
    \raisebox{-12pt}{
    \begin{tikzpicture}
        \draw [red, thick, decoration = {markings, mark=at position 0.6 with {\arrow[scale=1]{stealth}}}, postaction=decorate] (-1,0.5) -- node[below]{$\a$} (1,0.5) ;
        \draw (-1,0) rectangle (1,1);
    \end{tikzpicture}
    }
    =\raisebox{-26pt}{
    \begin{tikzpicture}
        \draw [red, thick, decoration = {markings, mark=at position 0.6 with {\arrow[scale=1]{stealth}}}, postaction=decorate] (0.5,0) -- node[left]{$\a$} (0.5,2) ;
        \draw (0,0) rectangle (1,2);
    \end{tikzpicture}
    }=Z^{\a}[\tilde{q}]\, , 
\end{align}
 where $\tilde{q}$ is obtained under a modular $S$-transformation $\tau \to -\frac{1}{\tau}$ and that, in the limit in which we work here, $L\gg T$, hence $\tilde{q} \to 0$, 
 \beq
 \frac{Z[q,\a]}{Z[q,\id]} \sim \tilde{q}^{\Delta_1^{\a}}\to 0\, ,
 \eeq
 where  $\Delta^{\a}_1>0$ is the scaling dimension of the vacuum corresponding to the vacuum of the $\a$-twisted sector\footnote{and taking $\Delta^{\id}_{1}=0$}. 

Generically, when a topological operator is defined along the whole space at a fixed time, it amounts to a conserved operator that acts on the Hilbert space $\cH$. If the topological operator is inserted in the time direction and localized in one of the spatial directions, the topological operator is a defect that modifies the quantization. The modified quantization gives rise to a twisted Hilbert space \cite{Hegde:2021}. We will not consider the later insertions in this work.

\section{Averages over Categories and Relative Entropy as Entropic order parameter}
\label{sec:averages}
In this section we introduce a generalization of averages over groups for the case of modular fusion categories (MFC) in RCFT. We first review the case of Abelian groups and then we introduce the generalization. Then, following \cite{Casini:2020}, the relative entropy between density matrices and their averages over a MFC, is proposed as an entropic order parameter.

\subsection{Group and Category averages}
In order to fix concepts, let us first recall the \emph{symmetrization} of $\rho$ as the average over a discrete Abelian group $G$ of the transformed density matrix $g\, \rho\, g^{-1}$, that is,
\begin{equation}\label{eq:def-symmetrized}
    \rho_{G}= \frac{1}{|G|}\sum_{g \in G}  \, g\, \rho\,  g^{-1},
\end{equation}
where $|G|$ is the normalized measure of the group, that can be generalized to generic compact Lie groups. Eq \eqref{eq:def-symmetrized} implements a map between density matrices known as  conditional expectation \cite{Petz:2004, Casini:2020}. In essence, this map extracts the portion of $\rho$ that remains unchanged under the action of the group $G$. The density matrix $\rho_{G}$ is by construction symmetric under $G$ and has trace one. Note that $\rho$ is symmetric if and only if $\rho=\rho_{G}$. Therefore, comparing the two states $\rho$ and $\rho_G$ would lead naturally to probing (spontaneous or explicit) symmetry-breaking \cite{Casini:2020}.

Here, we introduce the average over a category that can be naturally defined as
\begin{equation}
    \rho_{\cC}= \frac{1}{\N}\sum_{\a \in \cC}  \, \a\, \rho\, \bar{\a}\, ,
    \label{eq:cat_average}
\end{equation}
where $\bar{\a}$ is the conjugate of $\a$, that is $\a \times \bar{\a} = \id \oplus ...$. The constant $\N$ is fixed imposing that $ \rho_{\C}$ is a density matrix,
 \beq
 {\rm Tr}  \rho_{\C}  = 1 \Longrightarrow  \N = \sum_{\a,  \b \in \C}  N_{\a \bar{\a}}^\b {\rm Tr}[\rho\, \b]
 \label{2}
 \eeq
 For group-like invertible symmetries we will have
 \beq
 N_{\a \bar{\a}}^\b = \delta_{\b \id}  \Longrightarrow \N =  \sum_{\a \in \C}  {\rm Tr} [\rho\id]  = |\C |
 \label{3}
 \eeq
 and we will recover the standard definition of $\rho_{G}$. However, unlike the case of a group, this map generally does not implement a conditional expectation.

\subsection{Relative Entropy as an entropic order parameter}
The relative entropy between two reduced density matrices $\rho, \sigma$ is given by
\begin{equation}
S( \rho|| \sigma)=  {\rm Tr}\; \rho\log \rho-{\rm Tr}\; \rho\log \sigma\, ,
\end{equation}
that can be derived from
\begin{align} 
S( \rho|| \sigma) &= \lim_{n \to 1} \frac{1}{n-1} \log \left(\frac{{\rm Tr} [\rho^n]} {{\rm Tr} [\rho\, \sigma^{n-1}]} \right)\, . 
\end{align} 

Relevant properties of this quantity can be found in \cite{Petz:2004}. The relative entropy can be operationally interpreted as follows: given $\sigma$, the probability $p$ of mistaking $\rho$ for $\sigma$ after $N$ well-designed experimental measurements decays with $N$ as $p\sim e^{-N\, S( \rho|| \sigma) }$, which endorses the relative entropy as an experimentally accessible quantity.

Here, we  propose entropic order parameters that allow us to better understand the symmetry breaking of categorical symmetries in 2D CFT \cite{Bhardwaj:2023}. Following \cite{Casini:2020}, we investigate the behavior of the relative entropy $S(\rho_{G}|| \rho)$ (resp. $S(\rho_{\cC}|| \rho)$) between the ground states $\rho$ of the theory and their averaged versions $\rho_G$ (resp. $\rho_{\cC}$). It is expected that $S(\rho_{G}|| \rho)$ (resp. $S(\rho_{\cC}|| \rho)$) explicitly depend on the vacuum expectation values of the TDLs implementing the symmetries under consideration \cite{Casini:2020}.
For the group-like invertible symmetries, it is known that the relative entropy vanishes if the state $\rho$ is invariant under the action of the symmetry, i.e., $\rho$ is an unbroken symmetry vacuum, and then the two states that the relative entropy compares are identical. Nevertheless, if the group-like symmetry is broken, the relative entropy between these states must be necessarily non-zero. Hence, the usefulness of this concept as an order parameter in those cases.

Our aim here is to extend this program to the case of symmetries given by modular fusion categories in RCFTs.

\section{Relative Entropy in CFTs with Categorical Symmetries}
\label{sec:rel_ent_cat}
Here we compute the relative entropy between the category-averaged density matrix $\rho_{\C}$ and $\rho$ defined by
 \beq
  S(\rho_{\C}||\rho)=\lim_{n \to 1}\, \frac{1}{n-1} \log\, \left[\frac{{\rm Tr}\, [\rho_{\C}^n]}{{\rm Tr}\, \left[\rho_{\C}\, \rho^{n-1}\right]}\right]\, .
  \label{36}
  \eeq
 We focus first on the term ${\rm Tr}\, \left[\rho_{\C}\, \rho^{n-1}\right]$.  We note that all $\a \in \cC$  are topological and therefore, they are free to move along the Euclidean time direction between the sheets of the replicated manifold under continuous deformations. Indeed, these movements allow to fuse them. With this, one may write the term of interest as
 \begin{align}\label{37}
 {\rm Tr}\, \left[\rho_{\C}\, \rho^{n-1}\right] &= \frac{1}{\N}\sum_{\a \in \C}\, {\rm Tr}\, \left[\a \rho \bar{\a} \rho^{n-1}\right] = \frac{1}{\N}\sum_{\a \in \C}\, {\rm Tr}\, \left[\rho^n \a \bar{\a} \right] \\ \nonumber
 & = \frac{1}{\N}\sum_{\a, \b \in \C}\, N_{\a \bar{\a}}^{\b} {\rm Tr}\, \left[\rho^n \b \right] = \frac{1}{\N}\sum_{\a, \b , \c \in \C}\, \frac{(S_{\a \c})^2}{S_{1 \c}} S_{\b \c} {\rm Tr}\, \left[\rho^n \b \right] \\ \nonumber
 & = \frac{1}{\N}\sum_{\b , \c \in \C}\,\left(\sum_{\a \in \C} (S_{\a \c})^2 \right) \frac{S_{\b \c}}{S_{1 \c}}\,   {\rm Tr}\, \left[\rho^n \b \right] = \frac{1}{\N}\sum_{\b , \c \in \C}\, \frac{S_{\b \c}}{S_{1 \c}}\,   {\rm Tr}\, \left[\rho^n \b \right]\, ,
 \end{align}
where we have used the Verlinde formula and the fact that $\left(\sum_{\a \in \C} (S_{\a \b})^2 \right)=1$.

As we are interested in exploring the asymptotic limit $\tilde{q}\to 0$, it is interesting to write the above result as
 \begin{align}\label{38}
 {\rm Tr}\, \left(\rho_{\C}\rho^{n-1}\right) & = \frac{1}{\N}\left(\sum_{\c \in \C}  {\rm Tr}\, \left[\rho^n\right] + \sum_{\b\neq 1 , \c \in \C}\, \frac{S_{\b \c}}{S_{1 \c}}\,   {\rm Tr}\, \left[\rho^n \b \right]\right)\\ \nonumber
 &= \frac{|\C|}{\N}{\rm Tr}\, \left[\rho^n\right] + \frac{1}{\N}\sum_{\b\neq 1 , \c \in \C}\, \frac{S_{\b \c}}{S_{1 \c}}\,   {\rm Tr}\, \left[\rho^n \b \right]\\ \nonumber
 &=\frac{|\C|}{\N}{\rm Tr}\, \left[\rho^n\right]\left(1 + \frac{1}{|\C|}\sum_{\b\neq 1 , \c \in \C}\, \frac{S_{\b \c}}{S_{1 \c}}\,  \frac{ {\rm Tr}\, \left[\rho^n \b \right]}{{\rm Tr}\, \left[\rho^n\right]}\right)\, ,
 \end{align}
  because, as commented above, in the limit $\tilde{q}\to 0$, 
\beq\label{39}
 \frac{ {\rm Tr}\, \left[\rho^n \b \right]}{{\rm Tr}\, \left[\rho^n\right]} \approx  \tilde{q}^{\Delta^1_{\b}/n}\to 0\, ,
 \eeq
 with $\Delta^1_{\b}$  the scaling dimension of the vacuum corresponding to the vacuum of the $\b$-twisted sector. Therefore, the quantity
\beq\label{40}
  \Delta_n[\C]=\frac{1}{|\C|}\sum_{\b\neq 1 , \c \in \C}\, \frac{S_{\b \c}}{S_{1 \c}}\,  \frac{ {\rm Tr}\, \left[\rho^n \b \right]}{{\rm Tr}\, \left[\rho^n\right]}\,,
  \eeq
 is subleading and only appears  in the fusion category case, as in the group like case, $\b$ can only be the identity in $N_{\a\bar{\a}}^{\b}$ appearing in \eqref{37}, that is, $\Delta_n[\C]$ identically vanishes for the group-like case.  Now, by writing
  \beq\label{41}
  {\rm Tr}\, \left(\rho_{ \C}\rho^{n-1}\right) = {\rm Tr}\, \left[\rho^n\right]\left[\frac{|\C|}{\N}\, \left(1 + \Delta_n[\C]\right)\, \right]\, ,
 \eeq 
 one may express the relative entropy as
 \begin{align}\label{42}
 S(\rho_{\C}||\rho)&= \lim_{n \to 1}\, \frac{1}{n-1} \log\, \left[\frac{{\rm Tr}\, \left[\rho_{\C}^n\right]}{{\rm Tr}\, \left[\rho^n\right]\left[\frac{|\C|}{\N}\, \left(1 + \Delta_n[\C]\right)\, \right]}\right]\\ \nonumber 
 &= \lim_{n \to 1}\, \frac{1}{n-1} \left(\log\, \frac{{\rm Tr}\, \left[\rho_{\C}^n\right]}{ {\rm Tr}\, \left[\rho^n\right]}-\log\left[\frac{|\C|}{\N}\, \left(1 + \Delta_n[\C]\right)\right]\right) \\ \nonumber
 &= \lim_{n \to 1}\left(-\Delta S_n + \frac{1}{1-n}\,\log\left[\frac{|\C|}{\N}\, \left(1 + \Delta_n[\C]\right)\right]\right) \, .
 \end{align}
 
Noting that
 \begin{align}\label{43}
\N &= \sum_{\a, \b \in \C}  N_{\a \a}^\b {\rm Tr}[\rho\, \b]  = \sum_{\a, \b \in \C} \sum_\c \frac{ S_{\a  \c} S_{\a \c} S_{\b \c}}{S_{\id \c}} {\rm Tr}  [ \rho\, \b] \\ \nonumber
&=  \sum_{\b, \c } \frac{S_{\b \c}}{S_{1 \c}} \, {\rm Tr}  [ \rho \b] = 
 |\C|\, {\rm Tr}  [\rho]+ \sum_{\b \neq 1, \c } \frac{S_{\b \c}}{S_{1 \c}} \, {\rm Tr}  [ \rho \b]\\ \nonumber
 & = |\C|\, {\rm Tr}  [ \rho]\left(1 + \frac{1}{|\C|}\sum_{\b \neq 1, \c } \frac{S_{\b \c}}{S_{1 \c}} \, \frac{{\rm Tr}  [ \rho \b]}{{\rm Tr}  [ \rho]}\right) =  |\C|\,\left(1 + \Delta_1[\C]\right)\, ,
 \end{align}
where we have used that, for diagonal RCFTs, $\bar{\a}=\a$ and we have taken the definition of $\Delta_1[\C]$ from \eqref{45}. One may finally write 
\beq\label{44}
 S(\rho_{\C}||\rho) =\lim_{n \to 1}\left(-\Delta S_n + \frac{1}{1-n}\,\log\,\frac{ \left(1 + \Delta_n[\C]\right)}{\left(1 + \Delta_1[\C]\right) }\right)\, ,
\eeq 
where $\Delta S_n$ is given by,
\beq
 \Delta S_n =  \frac{1}{1-n} \log \frac{   {\rm Tr}  [\rho_{\C}^n]}{  {\rm Tr} [\rho^n]} \, .
 \label{45}
\eeq
This quantity, when restricted to a subsystem $A$ of the theory, is known as \emph{entanglement asymmetry}, quantifying the degree of symmetry breaking in a region of an extended quantum system \cite{Ares:2022, Fossati:2024}.

 We note that while for the group-like case, where $\Delta_n[\C]=0$ exactly vanishes and thus, $S(\rho_{A, \C}||\rho_A) = -\Delta S_1$, for the category case, at leading order in $\tilde{q} \to 0$ we have
 \beq\label{46}
 S(\rho_{\C}||\rho) \sim -\Delta S_1\, .
 \eeq
 
\paragraph{Computation of $\Delta S_n$.}
Using the results above, we are now prompted to compute $\Delta S_n$. Therefore, by inserting \eqref{eq:cat_average} into \eqref{45}, one obtains
 \beq
 \Delta S_n =  \frac{1}{1-n} \log  \frac{1}{\N^n}  \sum_{\a_1, \dots, \a_n \in \C}  \frac{    {\rm Tr} \left[   \rho\,  \bar{\a}_1 \a_2 \rho \bar{\a}_2  \dots  \bar{\a}_{n-1} \a_n \rho \bar{\a}_n  \a_1  \right]  }{  {\rm Tr}  [\rho^n]} \, .
 \label{46}
 \eeq
With this, the multi-charged moments can be defined as
 \beq
 Z[q^n, \aa] =  {\rm Tr} \left[   \rho\,  \bar{\a}_1 \a_2 \rho \bar{\a}_2  \dots  \bar{\a}_{n-1} \a_n \rho \bar{\a}_n  \a_1  \right]
 \label{7}
 \eeq
 where $\aa \in \cC^n$ is the $n$-tuple defined by $\aa = (\a_1, \dots, \a_n)$, and graphically reads as
\begin{equation}
Z[q^n, \aa] \equiv Z[q^n,\a_1, \cdots,\a_n]=
\vcenter{\hbox{\begin{tikzpicture}
\draw (-1,-2.5) rectangle (1,0.5);

\draw [red, thick, -stealth]  (-1,0.25) -- node(A)[below] {{\tiny $\a_1$}} (1,0.25);
\draw [red, thick, -stealth]  (-1,0.-0.5) -- node(A)[below] {{\tiny $\a_2$}} (1,0.-0.5);
\draw [red, thick, -stealth]  (-1,0.-2.0) -- node(A)[below] {{\tiny $\a_n$}} (1,-2.0);
\end{tikzpicture}}}
\, .
\end{equation}
 
Given this, now Eq. \eqref{7} can be written as 
 \beq
 \Delta S_n =  \frac{1}{1-n} \log  \frac{1}{\N^n}  \sum_{\aa \in \C^n}  \frac{   Z[q^n, \aa]  }{   Z[ q^n] } \, .
 \label{8}
 \eeq

As commented above, the TDL $\a \in \cC$ are  free to move up and down along the (Euclidean) time direction of the replicated manifold under continuous transformations, and this allows to fuse them. Using this again, one can write,
 \beq
 Z[q^n, \aa] =  {\rm Tr} \left[   \rho^n \bar{\a}_1 \a_2 \bar{\a}_2  \dots  \bar{\a}_{n-1} \a_n  \bar{\a}_n  \a_1  \right]\, .
 \label{10}
 \eeq
 Now, one may use the fusion algebra $\a \times \b = \sum_{\c}  N_{\a \b}^\c  \c$\,  to, assuming that $\bar{\a} =  \a, \; \forall \a \in \C$,  write, 
\beq
 {\a}_1 \a_2 {\a}_2  \a_3 {\a}_3   \dots  {\a}_{n-1}  {\a}_{n-1} \a_n  {\a}_n  \a_1  =  \sum_{\b's, \c's } N_{\a_1 \a_2}^{\b_1}    N_{\b_1 \a_2}^{\c_1} 
   N_{\c_1 \a_3}^{\b_2}     N_{\b_2 \a_3}^{\c_2}    N_{\c_2 \a_4}^{\b_3} N_{\b_3 \a_4}^{\c_3} \dots  N_{\c_{n-2} \a_n}^{\b_{n-1}}  
   N_{\b_{n-1} \a_n}^{\c_{n-1}} N_{\c_{n-1} \a_1}^{\d} \, \d
 \label{13}
 \eeq

Using the Verlinde formula (Eq. \eqref{eq:verlinde}) and general properties of the $S$-matrix, we obtain (see Appendix)
  \beq
  \sum_{\aa} 
 Z[ q^n, \aa] =  \sum_{\b\,  \in\cC}  \sum_{\c\,  \in\cC} \frac{1}{ S_{\id \c}^{2n-1}}\, S_{\b \c}\,  {\rm Tr} [\rho^n\,  \b]\, , 
 \label{19}
 \eeq

that can be represented as
\begin{equation}
 \sum_{\a_1, \cdots \a_n}
\vcenter{\hbox{\begin{tikzpicture}
\draw (-1,-2.5) rectangle (1,0.5);

\draw [red, thick, -stealth]  (-1,0.25) -- node(A)[below] {{\tiny $\a_1$}} (1,0.25);
\draw [red, thick, -stealth]  (-1,0.-0.5) -- node(A)[below] {{\tiny $\a_2$}} (1,0.-0.5);
\draw [red, thick, -stealth]  (-1,0.-2.0) -- node(A)[below] {{\tiny $\a_n$}} (1,-2.0);
\end{tikzpicture}}}=\sum_{\b,\, \c  \in\cC} \frac{1}{ S_{\id \c}^{2n-1}}\, S_{\b \c}\, 
\vcenter{\hbox{\begin{tikzpicture}
\draw (-1,-2.5) rectangle (1,0.5);
\draw [blue, thick, -stealth]  (-1,0.-1.0) -- node(A)[below] {{\tiny $\b$}} (1,0.-1.0);
\end{tikzpicture}}}
\, ,
\end{equation}

 Finally, $\Delta S_n$ in \eqref{45} can be written as
 \barray
 \Delta  S_n  = 
 \label{22} 
\frac{1}{1-n} \log \frac{1}{ {\rm Tr}[ \rho^n] }  
 \frac{ \sum_{\b , \c}   \frac{ 1}{ S_{\id \c}^{2n-1}}\, S_{\b \c}\,   {\rm Tr}[\rho^n \b]  }{
\left(  \sum_{\b, \c } \frac{S_{\b \c}}{S_{\id \c}} {\rm Tr}[ \rho\b] \right)^n} \, . 
 \earray 

\paragraph{Asymptotic limit and examples.} 
We provide here the asymptotic limit $\tilde{q}\to 0$ for $\Delta  S_n $ in Eq. \eqref{22}. This amounts to taking only the terms related to $\b=\id$ in the sums above, from which one can derive that
 \beq 
 \Delta S_n \sim \frac{1}{1-n} \log \frac{1}{{\rm Tr}[\rho^n]} \frac{\sum_{\c} \frac{1}{S_{\id c}^{2(n-1)}} {\rm Tr} [\rho^n]}{\left(\sum_{\c} {\rm Tr} [\rho\,  \id]\right)^n}= \frac{1}{1-n} \log \frac{1}{ |\C|^n} \sum_{\c \in \cC} S_{\id \c}^{2 (1-n)} \, .
 \label{26}
 \eeq
 In limit $n \rightarrow 1$ one finds
 
 \beq
  \Delta S_1  = \log |\C| +  \frac{2}{|\C|} \sum_{\c \in \cC} \log S_{\id \c} 
  \label{27}
  \eeq
  As it is said above, it would be expected that the relative entropy is related to the vacuum expectation values of the TDL implementing the symmetry under consideration. To see this, let us recall the definition of  quantum dimension, 
  \beq
  d_\c = \frac{ S_{\id \c}}{S_{\id \id}}
  \label{28}
  \eeq
 so we can write \eqref{27} as
 \beq
  \Delta S_1  = \log |\C| +  2 \log S_{\id \id}  +  \frac{2}{|\C|} \sum_{\c \in \cC} \log d_{\c} 
  \label{29}
  \eeq
  On the other hand 
  \beq
1=   \sum_{\c \in \cC} S_{\id \c} S_{\id \c} = S_{\id \id}^2 \sum_{\c \in \cC} d_\c^2 
\label{30}
\eeq
which implies
\beq 
S_{\id \id } = \frac{1}{\D}, \qquad \D  = \sqrt{ \sum_{\c \in \cC} d_\c^2}
\label{31}
\eeq
where $\D$ is the total quantum dimension of the modular tensor category. 
Replacing \eqref{31} into \eqref{29} gives,
 \beq
  \Delta S_1  = \log |\C| -  2 \log \D   +  \frac{2}{|\C|} \sum_{\c \in \cC} \log d_{\c} \, .
  \label{32}
  \eeq

Finally, we express $S(\rho_{\C}||\rho)$ in a more illustrative form  by observing that the equation above can be written as
 
 \beq
 -\Delta S_1 = \log \left( \frac{1}{|\C|} \sum_{\a \in \cC}  d_{\a}^2 \right) - \frac{1}{|\C|}\sum_{\a \in \cC} \log d_{\a}^2\, ,
 \eeq
 and therefore, according to our result in Eq \eqref{45}, at leading order for categories, and exactly for groups,
 
 \beq\label{eq:result}
 S(\rho_{\C}||\rho) = \log \left(\frac{1}{|\C|} \sum_{\a \in \cC} d_{\a}^2\right)- \frac{1}{|\C|}\sum_{\a \in \cC} \log d_{\a}^2\, ,
 \eeq
which implies that  $S(\rho_{\C}||\rho) \geq 0$ always. To see where this property comes from, we may rewrite the relative entropy as:
\begin{equation}
    S(\rho_\C || \rho) = \log  \frac{\left(\sum_{\a} d_{\a}^2\right) /|\cC|}{\sqrt[|\cC|]{\left(\prod_{\a} d_\a^2\right)}} \, .
\end{equation}

We notice that this is the quotient between the arithmetic mean and the geometric mean of $d_\a^2$. As $d_\a^2 \in \mathbb{R}^+$, we can invoke the AM-GM inequality \cite{Sedrakyan2018}, resulting in the desired property $S(\rho_\C || \rho ) \geq 0$. Moreover, the inequality only saturates if all the elements in the mean values are equal to each other. In our case, we always must have an element satifying $d_\id = 1$, i.e. the identity. This means that $S(\rho_\c || \rho) = 0$ if and only if $d_\a = 1 \:\: \forall \a$, that is, just for group-like symmetries.  This result is interesting as it shows that,
 \begin{enumerate}
 \item At leading order, the relative entropy $S(\rho_{\C}||\rho)$ only depends on the expectation values of the TDLs or quantum dimensions.
 \item As commented above, for groups, in which every symmetry operator is implemented by a TDL $\a$ with $d_{\a}=1$, $S(\rho_{G}||\rho) = 0$. As an example, for the $G=\mathbb{Z}_2$ group, we have two invertible TDLs $\lbrace\id, \eta\rbrace$ with $\eta\times \eta=\id$  and,
  \beq
   d_\id = 1, \quad  d_{\eta} = 1 
  \eeq
  Thus,
  \beq
  S(\rho_{G}||\rho) = -\Delta S_1  = 0\, .
  \eeq
 \item For systems with non-invertible symmetries represented by a fusion category $\C$, there are some TDL's $\a$  with $d_{\a}>1$, and therefore, $S(\rho_{\C}||\rho) > 0$.
 As an example, for the Fibonacci category $\cC_{\rm Fib}$ we have two TDLs $\lbrace\id, W\rbrace$ with $W\times W=\id + W$ and,
  \beq
   d_\id = 1, \quad  d_W = \phi 
  \eeq
  where $\phi = (1 + \sqrt{5})/2$.   Thus,
  \beq
  S(\rho_{\C}||\rho) = -\Delta S_1  = 0.111572...
  \eeq
  
  \item At leading order, the relative entropy $S(\rho_{\cC}||\rho)$ essentially counts the number of invariant configurations of the replica partition functions under the action of TDLs insertions, that is, for how many configurations $\aa \in \cC^n$, $Z[q^n,\aa] \sim Z[q^n,\id]$. It has been shown  that in spontaneously broken phases, the action of TDLs implementing symmetries on the vacua of the theory may be different depending on the phase \cite{Bhardwaj:2023, Bhardwaj:2023b}. In our setting, this would lead to a different counting on the configurations satisfying $Z[q^n,\aa] \sim Z[q^n,\id]$. This suggest that $S(\rho_{\cC}||\rho)$ may characterize distinct phases associated to the SSB of categorical-symmetries.
 \end{enumerate}

\section{Conclusions and Outlook}
\label{sec:conclusions}
We have studied the relative entropy between the ground states of  RCFTs representing the critical point of phase transitions between different symmetry-broken phases of theories with categorical symmetries and their \emph{symmetrized} versions. 

We find that at leading order, this relative entropy only depends on the expectation values of the TDLs or quantum dimensions of the topological operators implementing the symmetry. This dependence is such that, for group-like symmetries in which all the quantum dimensions implementing the symmetry are equal to one, this order parameter vanishes identically. On the other hand, in the case of categorical symmetries, we show that this order parameter is not zero. Our results suggest that our proposal can be used to characterize different broken phases of 2D-theories with categorical symmetries.
     
As said above, being the relative entropy an experimentally accessible quantity, we note that, in extended quantum systems such as the theories addressed here, taking a measure is concomitant of considering a specific subsystem. In this situation, it is thus sensible to extend our ideas by using tools from the theory of entanglement. The \emph{entanglement asymmetry}, commented above, has been recently introduced as a measure of symmetry breaking at a subsystem level.  It is worth investigating if our entropic order parameter for categorical symmetries can be extended to this framework, following the ideas in \cite{Das:2024}.

\section*{Acknowledgements}
We thank Arpit Das for many fruitful discussions on these and related topics. The work of P.S.-B. is supported by Fundaci\'on S\'eneca de la Regi\'on de Murcia, grant 21609/FPI/21. J.M.-V. thanks the financial support of Spanish Ministerio de Ciencia e Innovación PID2021-125700NAC22. G.-S. acknowledges financial support through the Spanish MINECO grant PID2021-127726NB-I00, the CSIC Research Platform on Quantum Technologies PTI-001 and the QUANTUM ENIA project Quantum Spain through the RTRP-Next
Generation within the framework of the Digital Spain 2026 Agenda.

\section*{Appendix}
In this appendix, we show how to derive Eq. \eqref{19} in the main text. Let us focus momentarily in the case $n=3$ where
  \beq
 {\a}_1 \a_2 {\a}_2   \a_3  {\a}_3  \a_1  =  \sum_{\b_1, \b_2 , \c_1, \c_2 }N_{\a_1 \a_2}^{\b_1}    N_{\b_1 \a_2}^{\c_1} 
   N_{\c_1 \a_3}^{\b_2}     N_{\b_2 \a_3}^{\c_2}     N_{\c_2 \a_1}^{\d} \, \d   
 \label{14}
 \eeq
Using  the Verlinde formula
 \beq
 N_{ij}^k = \sum_m  \frac{ S_{im} S_{jm} S_{km}}{ S_{\id m} }
 \eeq
 we obtain 
 \beq
\sum_{j,k}  N_{ij}^k   N_{kj}^l = \sum_{j,k}  \sum_{m, n}  \frac{ S_{im} S_{jm} S_{km}}{ S_{\id m} } \frac{ S_{kn} S_{jn} S_{lm}}{ S_{\id l} } =
\sum_m \frac{ S_{im} S_{lm}} { S_{\id m}^2} 
 \eeq
 where we used that $\sum_j S_{ij} S_{jk} = \delta_{ik}$ and $S_{ij} = S_{ji}$.   
 The sum over the labels in \eqref{14} gives
   \barray 
   \sum_{\a_1, \a_2, \a_3} 
 {\a}_1 \a_2 {\a}_2   \a_3  {\a}_3  \a_1  & = &  \sum_{\a_1, \c_1, \c_2}   \left( \sum_{\a_2, \b_1} N_{\a_1 \a_2}^{\b_1}     N_{\b_1 \a_2}^{\c_1}  \right) 
 \left( \sum_{\a_3, \b_2}   N_{\c_1 \a_3}^{\b_2}     N_{\b_2 \a_3}^{\c_2}   \right)   N_{\c_2 \a_1}^{\d} \, \d   
 \label{17} \\
 & = & \sum_{\a_1, \c_1, \c_2}   \sum_{m_1, m_2} \frac{ S_{\a_1 m_1} S_{\c_1 m_1}} { S^2_{\id m_1} } 
 \frac{ S_{\c_1 m_2} S_{\c_2 m_2}} { S^2_{\id m_2} } N_{\c_2 \a_1}^{\d} \, \d    \nonumber \\ 
  & = & \sum_{\a_1, \c_1, \c_2}   \sum_{m_1, m_2, m_3} \frac{ S_{\a_1 m_1} S_{\c_1 m_1}} { S^2_{\id m_1} } 
 \frac{ S_{\c_1 m_2} S_{\c_2 m_2}} { S^2_{\id m_2} } \frac{ S_{\c_2 m_3} S_{\a_1 m_3} S_{\d m_3}} { S_{\id m_3} } \, \d  \nonumber \\
  & = &   \sum_{m_1, m_2, m_3} \frac{ \delta_{m_1 m_2} \delta_{m_2 m_3} \delta_{m_3 m_1} S_{\d m_3}}{
  S^2_{\id m_1} S^2_{\id m_2} S_{\id m_3} } 
    \, \d  \nonumber  = \sum_m  \frac{ S_{\d m}}{ S_{\id m}^5} \, \d  \nonumber 
 \earray 
 For arbitrary $n$ it is easy to find the general results which reads as,
  \beq
  \sum_{\a_1, \dots, \a_n} 
 {\a}_1 \a_2 {\a}_2  \dots  {\a}_{n-1} \a_n  {\a}_n  \a_1  = \sum_m  \frac{ S_{\d m}}{ S_{\id m}^{2n-1}} \, \d\, ,
 \label{18}
 \eeq
and hece
  \beq
  \sum_{\aa} 
 Z[q^n, \aa] =  \sum_\d \sum_m  \frac{ S_{\d m}}{ S_{\id m}^{2n-1}}  {\rm Tr}(\rho^n \d) \, .
\eeq

\newpage

\bibliographystyle{utphys}

\begin{thebibliography}{99}
\bibitem{Gaiotto:2014}
D.~Gaiotto, A.~Kapustin, N.~Seiberg and B.~Willett,
``Generalized Global Symmetries,''
JHEP \textbf{02} (2015), 172
doi:10.1007/JHEP02(2015)172
[arXiv:1412.5148 [hep-th]].


\bibitem{McGreevy:2022}
J.~McGreevy,
``Generalized Symmetries in Condensed Matter,''
doi:10.1146/annurev-conmatphys-040721-021029
[arXiv:2204.03045 [cond-mat.str-el]].

\bibitem{Bhardwaj:2023}
L.~Bhardwaj, L.~E.~Bottini, D.~Pajer and S.~Schafer-Nameki,
``Categorical Landau Paradigm for Gapped Phases,''
[arXiv:2310.03786 [cond-mat.str-el]].

\bibitem{Bhardwaj:2023b}
L.~Bhardwaj, L.~E.~Bottini, D.~Pajer and S.~Sch\"afer-Nameki,
``Gapped Phases with Non-Invertible Symmetries: (1+1)d,''
[arXiv:2310.03784 [hep-th]]

\bibitem{Casini:2020}
H.~Casini, M.~Huerta, J.~M.~Magan and D.~Pontello,
``Entropic order parameters for the phases of QFT,''
JHEP \textbf{04} (2021), 277
doi:10.1007/JHEP04(2021)277
[arXiv:2008.11748 [hep-th]].

\bibitem{Petz:2004}
M.~Ohya, D.~Petz
"Quantum entropy and its use"  
Science and Business Media, New York U.S.A

\bibitem{Casini:2019}
H.~Casini, M.~Huerta, J.~M.~Mag\'an and D.~Pontello,
``Entanglement entropy and superselection sectors. Part I. Global symmetries,''
JHEP \textbf{02} (2020), 014
doi:10.1007/JHEP02(2020)014

\bibitem{Magan:2020}
J.~M.~Magan and D.~Pontello,
``Quantum Complementarity through Entropic Certainty Principles,''
Phys. Rev. A \textbf{103} (2021) no.1, 012211
doi:10.1103/PhysRevA.103.012211


\bibitem{Chang:2018}
C.~M.~Chang, Y.~H.~Lin, S.~H.~Shao, Y.~Wang and X.~Yin,
``Topological Defect Lines and Renormalization Group Flows in Two Dimensions,''
JHEP \textbf{01} (2019), 026
doi:10.1007/JHEP01(2019)026
[arXiv:1802.04445 [hep-th]].


\bibitem{Verlinde:1988}
E.~P.~Verlinde,
`Fusion Rules and Modular Transformations in 2D Conformal Field Theory,''
Nucl. Phys. B \textbf{300} (1988), 360-376
doi:10.1016/0550-3213(88)90603-7

\bibitem{Hegde:2021}
S.~Hegde and D.~P.~Jatkar,
``Defect partition function from TDLs in commutant pairs,''
Mod. Phys. Lett. A \textbf{37} (2022) no.29, 2250193
doi:10.1142/S0217732322501930



 \bibitem{Lin:2022}
Y.~H.~Lin, M.~Okada, S.~Seifnashri and Y.~Tachikawa,
``Asymptotic density of states in 2d CFTs with non-invertible symmetries,''
JHEP \textbf{03} (2023), 094
doi:10.1007/JHEP03(2023)094
[arXiv:2208.05495 [hep-th]].

\bibitem{Ares:2022}
F.~Ares, S.~Murciano and P.~Calabrese,
``Entanglement asymmetry as a probe of symmetry breaking,''
Nature Commun. \textbf{14} (2023) no.1, 2036
doi:10.1038/s41467-023-37747-8
[arXiv:2207.14693 [cond-mat.stat-mech]].

\bibitem{Fossati:2024}
M.~Fossati, F.~Ares, J.~Dubail and P.~Calabrese,
``Entanglement asymmetry in CFT and its relation to non-topological defects,''
JHEP \textbf{05} (2024), 059
doi:10.1007/JHEP05(2024)059

\bibitem{Sedrakyan2018}
Sedrakyan, Hayk; Sedrakyan, Nairi (2018),
Sedrakyan, Hayk; Sedrakyan, Nairi (eds.),
"The HM-GM-AM-QM Inequalities",
\textit{Algebraic Inequalities}, 
Problem Books in Mathematics,
Cham: Springer International Publishing, 
p. 21,

\bibitem{Das:2024}
A.~Das, J.~Molina-Vilaplana and P.~Saura-Bastida,
``Generalized Symmetry Resolution of Entanglement in CFT for Twisted and Anyonic sectors,''
[arXiv:2409.02162 [hep-th]].

\end{thebibliography}

\end{document}